\documentclass[final, 11pt,oneside]{IEEEtran}

\usepackage{cite}
\usepackage{graphicx,enumerate,enumitem} 
\usepackage{subcaption}

\usepackage{amsfonts,mathtools,amsmath,amssymb,color,array,esvect,bm}

\begin{document}

\title{
Semantics-Empowered Communication \\for Networked Intelligent Systems }
\author{
Marios Kountouris and Nikolaos Pappas
\thanks{Marios Kountouris is with the Communication Systems Department, EURECOM, Sophia Antipolis 06904, France (email: marios.kountouris@eurecom.fr). Nikolaos Pappas is with the Department of Science and Technology, Link\"{o}ping University, SE-60174 Norrk\"{o}ping, Sweden (email: nikolaos.pappas@liu.se). }
}

\maketitle

\begin{abstract}
Wireless connectivity has traditionally been regarded as an opaque data pipe carrying messages, whose context-dependent meaning and effectiveness have been ignored. 
Nevertheless, in emerging cyber-physical and autonomous networked systems, acquiring, processing, and sending excessive amounts of distributed real-time data, which ends up being stale or useless to the end user, will cause communication bottlenecks, increased latency, and safety issues.
We envision a communication paradigm shift, which makes the {\emph{Semantics of Information}}, i.e., the significance and usefulness of messages, the foundation of the communication process. 
This entails a goal-oriented unification of information generation, transmission, and reconstruction, by taking into account process dynamics, signal sparsity, data correlation, and semantic information attributes. 
We apply this structurally new, synergetic approach to a communication scenario where the destination is tasked with real-time source reconstruction for the purpose of remote actuation. Capitalizing on semantics-empowered sampling and communication policies, we show significant reduction in both reconstruction error and cost of actuation error, as well as in the number of uninformative samples generated. 
\end{abstract}

%========================================================================================================%
%========================================================================================================%
\section{Introduction}
Today’s communication technology offers a cornucopia of wireless connectivity solutions and is the foundation of our hyperconnected society and automated economy. The interconnection of myriad autonomous smart devices (robots, vehicles, drones, etc.) empowered with advanced sensing, computing, and learning capabilities, is forecast to generate a staggering amount of data (in the order of zettabytes). For example, data gathered by an autonomous car starts from 750 MB per second. A swarm of mobile robots may involve transmission of 1 GB aggregated data per second for target tracking or collaborative sensing.

In this expanding ecosystem, wireless networks are evolving to cater to emerging cyber-physical and mission-critical interactive systems, such as swarm robotics, self-driving cars, and smart Internet of Things (IoT). These systems call for reliable real-time communication, autonomous interactions, and automated decision making. Their successful operation entails the processing and the exchange of massive volumes of multimodal, often high dimensional, distributed data, in an efficient, effective and timely manner. Simply generating and communicating data traffic, which often ends up being outdated or irrelevant to the end user's application, will cause severe communication bottlenecks. These bottlenecks could inevitably jeopardize the proper functioning of wireless networks, leading to unnecessary network congestion, wasteful resource utilization, and excessive energy consumption.

\subsection{The End of Current Communication Paradigm?}
Virtually all today’s wireless systems are built upon fundamental principles of reliable communications over noisy channels, first developed in the \textit{locus classicus} of information theory \cite{Shannon48}. 
Despite various endeavors \cite{Sem1,Sem2,Sem3,Sem4,Sem5}, most existing communication paradigms are content-agnostic, in particular at the lower protocol layers where the significance and the effectiveness of transmitted messages have been set aside. 
The main objective has been to optimize conventional key performance indicators, such as throughput, delay, and packet loss; quality of service is usually provided through network over-provisioning and resource reservation control.

The dichotomy of information content and significance was a conceptual advance, which has been suitable for classical data communication targeting error-free high-speed data transmission. 
In sharp contrast, this approach comes short of meaningfully scaling and of supporting the needs of emerging networked intelligent systems and machine-type communication. Consider, for example, a larger number of autonomous mobile robots communicating with one another to reach timely consensus in the negotiation for collision avoidance. 
Achieving this goal is neither simply a question of understanding the throughput-reliability-delay tradeoff, nor of delivering streams of ``random” bits from one robot to another while maximizing throughput or minimizing delay.  
For a safe and successful operation, it is crucial to factor into the communication process the \textit{urgency} and the \textit{value} of messages provided by each robot. That way, transmissions are prioritized efficiently and the application demands are met with greater accuracy.

Recent work on status update systems (e.g., environmental monitoring, news reports, Web crawlers, etc.) and networked control systems has started addressing data prioritization issues. Therein, first steps towards importance-aware communication have harnessed the concepts of age of information (AoI) and its variants \cite{NowAoI,yates2020age, sunmodiano2019age,altman2019forever,AoII}, value of information (VoI) \cite{VoI_USSR,VoI,Molin,Kellerer,JIoT2020,GCW2019Alarms}, and quality of information \cite{QoI}. 
Several application-driven technologies have recently emerged for the upper layer network management and orchestration, including named-data networking \cite{NDN}, semantic-plane protocols \cite{Popovski}, zero-touch networks, as well as software-defined and intent-based networking. Most of these networking paradigms rely on high-level abstractions and leverage machine learning to configure and continuously maintain the network in a desired state according to business intents. 

\subsection{Towards Goal-oriented Semantic Communication}
Looking beyond the aforementioned confined view of wireless connectivity, we place ourselves in a setting where communication is not an end in itself but a means to achieving specific {\emph{goals}}.
Our vision entails a communication paradigm shift to enable the generation and the timely provisioning of the appropriate information to the right processing point. 
This can be realized, in a nutshell, by making the \emph{Semantics of Information} the foundation of the entire communication process. 
Differently from its common use in linguistics, logic, or computer science (e.g., Semantic Web, databases, ontologies, etc.), semantics is employed here with its etymological meaning, that of significance. Semantics here is a measure of the usefulness of messages with respect to the goal of data exchange.
Concretely, we propose a semantics-empowered communication system, whose foundations entail a goal-oriented unification of data generation, information transmission, and usage. Our approach capitalizes on the largely untapped innate and contextual attributes (semantics) of information. That way, the entire communication process can be tailored to meet the networked applications' requirements, thus enabling to achieve specific goals.
Considering a simple actuation-oriented real-time reconstruction scenario, we show that this paradigm shift has the potential to entirely transforming several prevailing design principles. We showcase its potential to significantly reduce the number of uninformative data samples, the real-time reconstruction error and the cost of actuation error.

%========================================================================================================%
%========================================================================================================%
\section{Semantics-empowered Communication}\label{sec:SemComm}

\subsection{Defining the Semantics of Information}\label{sec:SoI}
A first natural question is how to define and measure the information semantics. We advocate for assessing and extracting the semantic value of data at three different granularity levels.

\subsubsection{Microscopic scale}
At the source level, semantics refers to the relative importance of different, often equiprobable, events, outcomes, or observations from a stochastic source of information or a process. For instance, these primary information sources could represent sensor measurement data, patterns of a physical phenomenon (e.g., vehicle's trajectory), or the state of a dynamical system. 
Imagine two equally rare events, occurring with very low probability ($p \ll 1 $), one of which carries a major safety risk while the other is just a peculiarity. Although they provide the same high amount of information ($-\log p$), the information conveyed by the first event is evidently of higher significance. This disparity in importance can be incorporated into key information measures (e.g., entropy rate, mutual information) and statistical similarity metrics (e.g., $f$-divergences) using weight functions. These functions may be context-dependent and could incorporate various temporal variations and spatial patterns in information utility. To its simplest form, we can define a context-dependent entropy defined as $H(P)=-\sum_{y\in\mathcal{Y}}\varphi(y)P(y)\log P(y)$, for a probability mass function $P$ on a discrete set $\mathcal{Y}$ and a function $\varphi(\cdot)$ that weights the different outcomes with respect to their utility for a specific goal. Semantics can also be captured using R\'enyi's information measures \cite{Renyi}, which are instrumental in assessing compressability \cite{CompTIT18}, sparsity, and trackability \cite{BacISIT2020} of stochastic processes, signal complexity \cite{BaraniukTIT01}, as well as information gain efficiency in decision making (e.g., robotic exploration, importance sampling, multi-goal reinforcement learning). 

\subsubsection{Mesoscopic scale}
At the link level, semantics of information is a composite nonlinear multivariate function of the vector of information attributes. These qualitative attributes of information can be either \textit{innate} (objective) or \textit{contextual} (subjective). The former are attributes inherent to information regardless of its use, i.e., they depend on the information generated by a source and on its transformations (e.g., compression). 
Representative innate attributes include \textit{freshness} or AoI, i.e., $A_t=t-\nu_t$ where $\nu_t$ is the generation time of the newest sample that has been delivered to the receiver by time instant $t$, and \textit{precision}, i.e., a measure of the degree of closeness of measured values to each other and of the reproducibility of the measurement. 
The latter are attributes that depend on the particular context or goal for which information is being used. The most relevant ones are \textit{timeliness}, i.e., the time instant by which information has to be available for use at a point of computation or decision making, and \textit{completeness}, an information relevance attribute that measures the difference between the information amount and the total information of the real world. For example, an image delivered by the network for remote monitoring has a certain freshness, field of view, and resolution (precision). Mission-critical applications impose stringent requirements on timeliness and availability of data. 

A widely used attribute is \textit{accuracy}, which describes the degree of correctness and can be perceived as both intrinsic and contextual. It is related to distortion, i.e., the distance between the measured or estimated value or state and the true value or state.
As stated above, information semantics can be formally defined as a composite function (or a functor in the general case dealing with categories and topological structures), where a context-dependent, cost-aware function is applied to a multidimensional function of information attributes. Formally, the semantics of information $\mathcal{S}_t = \nu(\psi(\mathcal{I},\mathcal{C}))$, where $\psi(\mathcal{I},\mathcal{C})$ is a multidimensional, non-linear function of the vector of innate ($\mathcal{I}$) and contextual ($\mathcal{C}$) attributes, and $\nu(\cdot)$ is a context-dependent, cost-aware function that maps qualitative information attributes to their application-dependent semantic value. For example, in a simple scenario, $\mathcal{S}_t = w_1\mathcal{A} + w_2\mathcal{T}$, i.e., a weighted sum of error-based accuracy $\mathcal{A}$ and timeliness $\mathcal{T}$, where $\mathcal{T}=e^{-\gamma A_t}$ (contextual) could in turn be an exponentially decreasing function of information aging (innate) and $\gamma$ a parameter capturing the latency sensitivity \cite{KostaTCOM2020}.

\emph{Information value dualism:} It is important to highlight that information may have a value \textit{per se}, in addition to its ``utilitarian", context-dependent value. For instance, the precision of a sensor measurement has an intrinsic value related to the quality of how accurately it represents a phenomenon, whereas this same measurement has different value depending on its context of use and the application requirements (e.g., whether it monitors temperature in a smart home or in a nuclear plant). In Section~\ref{sec:example}, we introduce the cost of actuation error, which may cast the reconstruction error to the context of the application.

\subsubsection{Macroscopic scale}
At the system level, semantics of information is related to the end-to-end, effective distortion and timing mismatch between information $X_{t_1}$ generated at a point or region in spacetime $(\vv{\bm{x}}_1,t_1)$ and its reconstructed or estimated version $\hat{X}_{t_2}$ at another point in spacetime $(\vv{\bm{x}}_2,t_2)$, factoring in all sources of variability and latency (sensing latency and accuracy, data gathering, transmission latency, decoding, processing, etc.). The spacetime coordinates represent an event. For example, the original information may represent the system state of the physical world at a certain area, while its estimation may represent the perceived information about this physical world at a remote unit (virtual world, digital twin). In many real-time networked systems, the objective is to provide the observer at the receiver side with an “instantaneous” and accurate estimate of the information generated at the transmitter side.
This calls for a relativistic information transfer theory that allows us to synchronize the state evolution at both communication ends. In other words, the evolution of reconstructed information is aligned to the temporal dynamics of the original information source, while at the same time the “closeness” between the estimated and the “true” state is maximized. Roughly speaking, for some functional $\psi(\cdot)$, we would like to minimize $\psi(\hat{X}_{t_2}, X_{t_1})$ (e.g., $\lVert \hat{X}_{t_2} - X_{t_1}\rVert^2\to 0$) while $\left|t_2 - t_1 \right|\to 0$. That could minimize the time duration a remote unit remains in an erroneous and/or time mismatched state, via necessary transformations compensating for the system timing “dilation” (to draw an analogy with relativistic clock synchronization).
This holds the promise to provide the theoretical foundations for applications targeting real-time experience, such as extended reality, tactile internet, and holographic communication.

%========================================================================================================%
\subsection{Semantics-empowered Communication Model}\label{sec:CommModel}
In this section, we present the envisioned semantic communication model. In sharp contrast to most prevailing communication systems that assume uncontrolled exogenous traffic arrivals, the communication process in our proposed architecture starts from information generation and data acquisition. 
This radical departure capitalizes on smart devices’ ability to control their traffic via semantic-aware active sampling, in which samples are generated at will or triggered based. 
Furthermore, the entire communication process extends up to goal-oriented signal reconstruction and information usage and exploitation. 

\begin{figure}[ht]
	\centering
	\includegraphics[width=1\columnwidth]{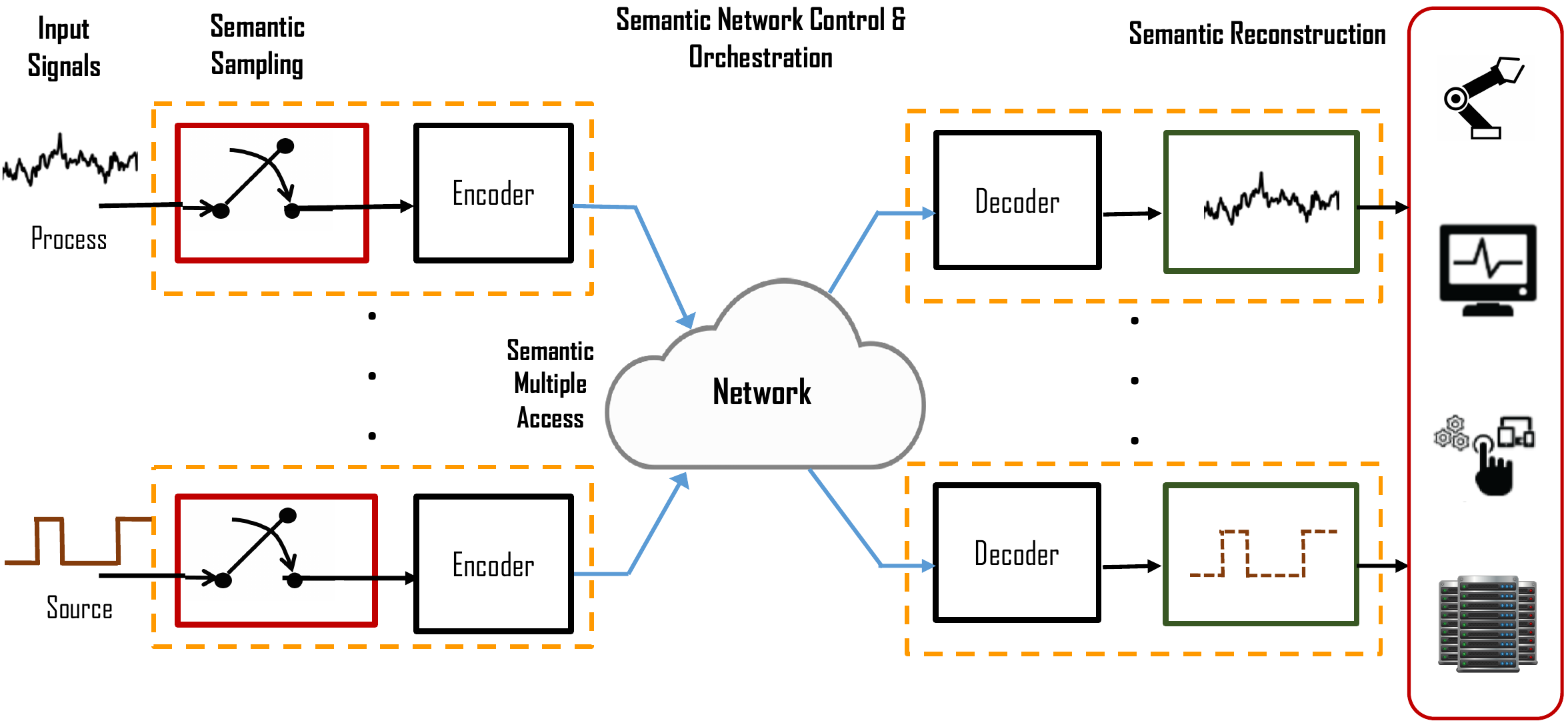} %\vspace{-0.2in}
	\caption{End-to-end goal-oriented semantic communication model.}
	\label{Fig:model}
\end{figure}  

The general end-to-end semantic communication model is depicted in Fig. \ref{Fig:model} and mainly includes the following building blocks.  
\begin{itemize}[leftmargin=*,align=left]
\item Multiple continuous or discrete time, possibly correlated, signals (stochastic processes) and information sources, which represent a time-varying real-world physical phenomenon in space, are observed by spatially distributed smart devices. In general, these devices may have heterogeneous sensing, computational and learning/inference capabilities.  
\item Smart devices access a shared communication medium to send data samples (e.g., observations, measurements, updates) to one or multiple destinations (e.g., fusion center, control unit). 
Samples are \textit{generated at will} using process-aware, non-uniform active sampling, according to the source variability (e.g., changes, innovation rate, autocorrelation, self-similarity), the communication characteristics, and the semantics-aware applications' requirements. That way, only the “most valuable and informative” samples are generated and prioritized for transmission.
\item Source samples could be preprocessed prior to being encoded and scheduled for transmission over noisy and delay/error-prone communication channels. This operation may include quantization, compression, and feature extraction, to name a few. Scheduling is performed according to semantic information value and priority,  extracted from data.
\item The input signals (sources) are finally reconstructed at the destinations from causally or non causally received samples to serve the application purpose, e.g., collision avoidance, remote state estimation, control and actuation, situation awareness, and learning model training, to name a few. In general, the reconstructed signals may alter the recipients' states and may initiate specific actions at the receiving ends (actionable intelligence).
\end{itemize}

%========================================================================================================%
%========================================================================================================%

\subsection{Joint sampling, communication, and reconstruction under real-time constraints}\label{sec:joint}
A fundamental element of the proposed communication paradigm is the cohesion of the entire process of information generation, transmission, and reconstruction, which has to be synergistically redesigned under the prism of semantic information. 
Let us highlight this with an example from networked robotics. A mobile robot generates and sends updates of a continuous stochastic process (e.g., a vehicle’s trajectory) to a remote tracking unit for real-time causal reconstruction or remote estimation \cite{Sun1,Sun2,GuoKost}. Conventional approaches decouple sampling from transmission, resulting in simple yet suboptimal solutions. Sampling is optimized based on signal’s changes, therefore samples might become stale before being successfully received. Transmission is optimized based on quality of service metrics (e.g., delay, rate, timeliness) ignoring the source variations; samples may be received on time but they contain no useful information or could even be misleading about the system's true state. This simple example reveals the structural links between sampling and communication, which are generally non-separable in semantic communication. This means that one cannot just take the best sampling policy, place it before the best communication scheme, and expect to get the best out of both. 
The key challenge is to develop a theory of optimal, semantics-aware joint active sampling, transmission, and  reconstruction of multidimensional signals, in particular under stringent timing constraints. 
This is of cardinal importance for enabling timely decision making and for efficiently meeting the requirements of real-time networked applications.

%========================================================================================================%
%========================================================================================================%
\section{An Illustrative Example}\label{sec:example}
\begin{figure}[ht]
	\centering
	\includegraphics[width=0.95\columnwidth]{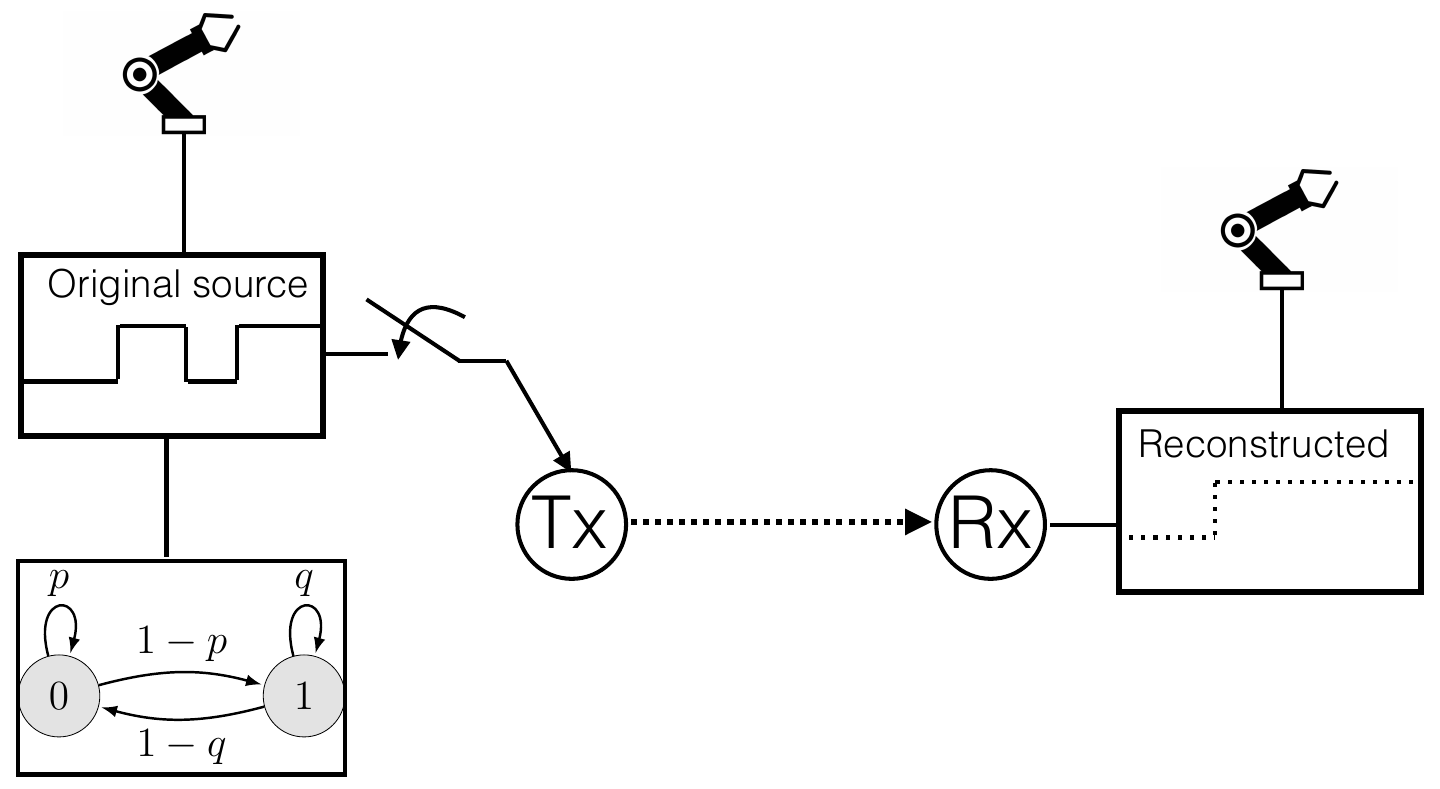}
	\caption{The setup for the illustrative example.}
	\label{fig:system}
\end{figure}

We consider an end-to-end communication system, in which a device monitors a two-state Markovian source. The source initiates certain actions to a robotic object at the transmitter side and the goal is to have a digital twin of that robotic object at the receiver side (Fig. \ref{fig:system}). 
We consider a slotted-time system in which the monitoring device samples the process and transmits updates on the source's status to a remote actuator. Status update packets are transmitted over a wireless erasure channel, in which realizations are independent and identically distributed over time slots. We consider two cases, one with low channel quality, in which the probability of successful transmission is $P_s = 0.4$, and one with high channel quality and success probability $P_s = 0.9$.
Real-time source reconstruction is performed at the end point, upon receipt of status updates, as a means to achieve the real-time actuation goal of the digital twin.

\subsection{Sampling and Transmission Policies}
We consider the following four policies for information generation and transmission. 

\textbf{Uniform.} In this source-agnostic policy, sampling is performed periodically, independently of the evolution of the source process. There could be thus several state transitions (changes) between two collected samples, especially for rapidly varying sources. If transmission fails, the most recently acquired measurement (sample) is communicated. 

\textbf{Age-aware.} In this policy, acquisition and transmission of a new sample is triggered by the receiver once the AoI reaches a given threshold. Note that this could also model the case where the transmitter knows the AoI at the receiver's side (using feedback acknowledging or not the receipt of sample). Since AoI can be viewed as a concrete, quantitative surrogate for semantics, the age-aware policy can be considered as a first, simple semantics-aware scheme. 
Whenever a transmission fails, the receiver tries to anticipate the update based on the statistics of the source process. In that case, the receiver, given its current state, tries to predict the next state based on the state transition probabilities, which are assumed to be known (or can be learnt after a period of time). 

\textbf{Semantics-aware.} This is a source change-triggered policy, i.e., sample generation is triggered at the transmitter side whenever a change at the state of the source is observed (since the previous sample). In a way, this can be seen as a VoI-aware sampling policy. 
Consider that in a given time slot, the source is in a certain state, in which it remains for a certain period of time. At the end of that period, the state changes, hence the transmitter generates and transmits a new status update sample. Notice that this policy takes into account only changes occurring and tracked at the source side (transmitter). 

\textbf{E2E Semantics.} In this end-to-end policy, sample acquisition is triggered whenever there is discrepancy between the states at the two communication ends. It extends the semantics-aware policy so that the amount of change is not solely measured at the source, but is tracked by the difference in state between the two ends. Let us clarify what is meant by difference here. Assume that in a given time slot, both source and destination are in the same state. Then, a change in the state at the source occurs in the next slot; hence a new sample is generated and transmitted. In case of erasure, the reconstructed source will remain in the previous state. In the next slot, the original source returns back to the state that was two time slots before. This means that no discrepancy exists now between the original and the reconstructed source, thus, there is no need for sending an update. 

\subsection{Metrics and Performance Evaluation}
Performing sampling and transmission at each slot could evidently provide the best result for the application. However, this approach does not scale; an excessive number of (not necessarily useful) samples is generated, which require a tremendous increase in communication resources for their transmission. Our semantic approach primary aims at reducing or even eliminating the generation of uninformative sample updates, thus improving network resource usage.

Performance is assessed using the following metrics of interest: \textit{real-time reconstruction error} and \textit{cost of actuation error}. The reconstruction error measures the discrepancy in real-time of the values between the original and the reconstructed source as time evolves. The cost of actuation error captures the significance of the error at the actuation point given the fact that some errors can be non-commutative and may have higher impact than others. We consider here two cases for the error occurrence: when the original source is in the first state, but the reconstructed source believes that is in the second state, the cost of actuation is low (e.g., set to one), while in the opposite case the cost is assumed to be relatively high (e.g., equal to five). The latter corresponds to a case where the penalty or the loss from taking a wrong action upon a misconceived system's state is high.
When the sources are in the same state, there is no actuation error.

\begin{figure}
	\centering
	\begin{subfigure}{0.32\textwidth} 
		\includegraphics[width=\textwidth]{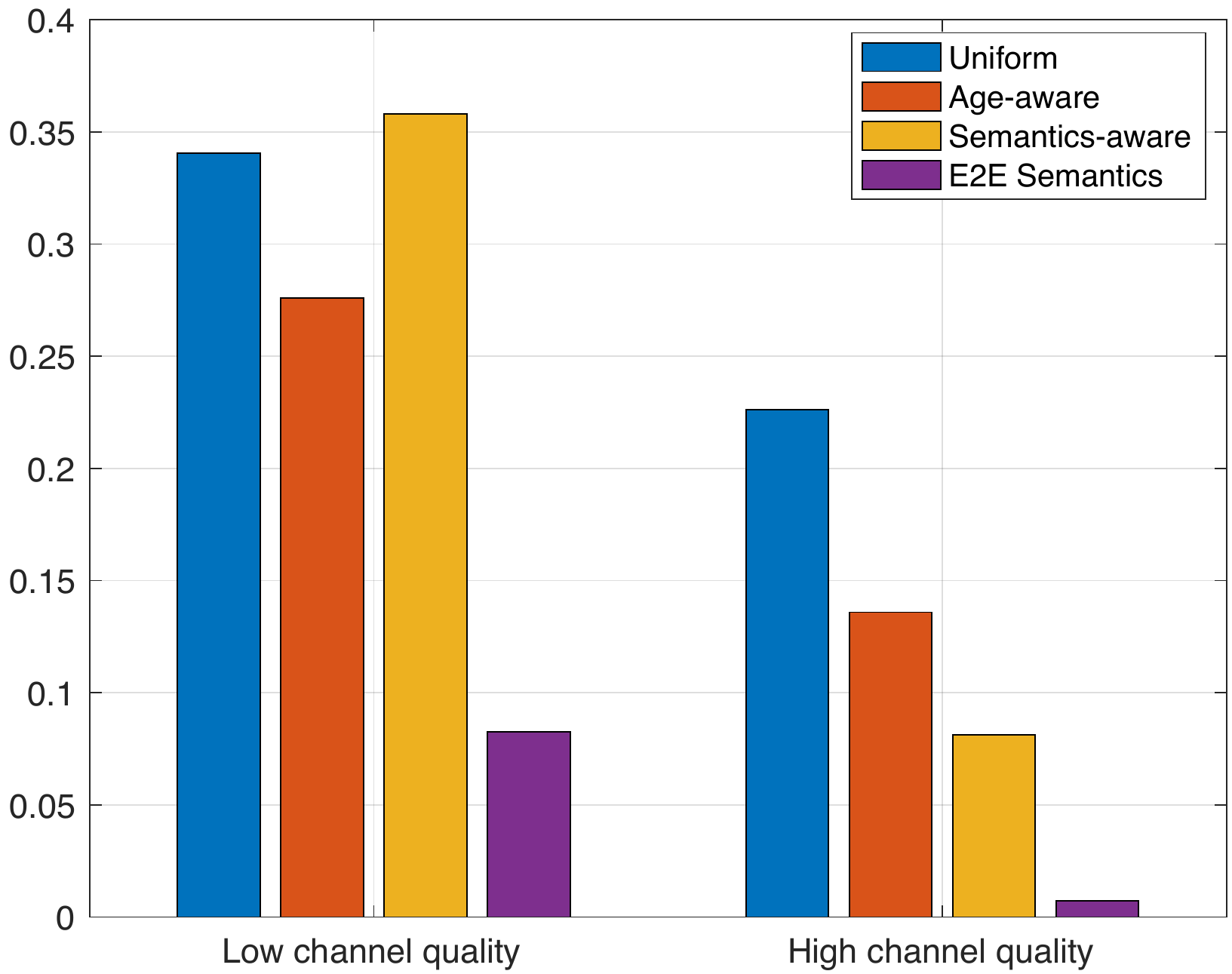}
		\caption{Real-time reconstruction error} 
		\label{fig:error_low_var}
	\end{subfigure}
	\vspace{1em} 
	\begin{subfigure}{0.32\textwidth} 
		\includegraphics[width=\textwidth]{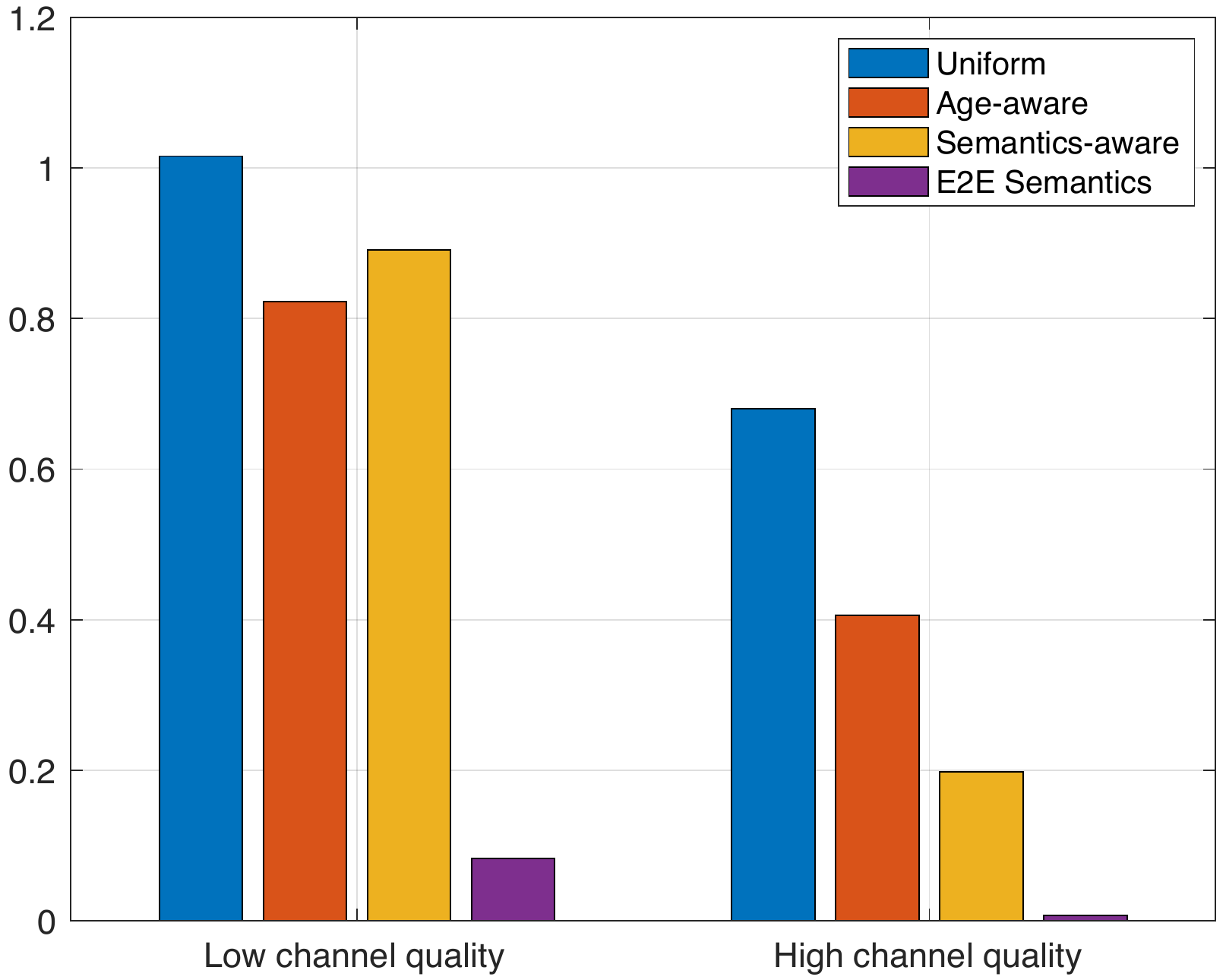}
		\caption{Cost of actuation error} 
		\label{fig:costacterror_low_var}
	\end{subfigure} \vspace{-0.2in}
	\caption{The case of a slowly varying source with $p=0.95$ and $q=0.9$.} 
	\label{fig:low_var}
\end{figure}

\begin{figure}
	\centering
	\begin{subfigure}{0.32\textwidth} 
		\includegraphics[width=\textwidth]{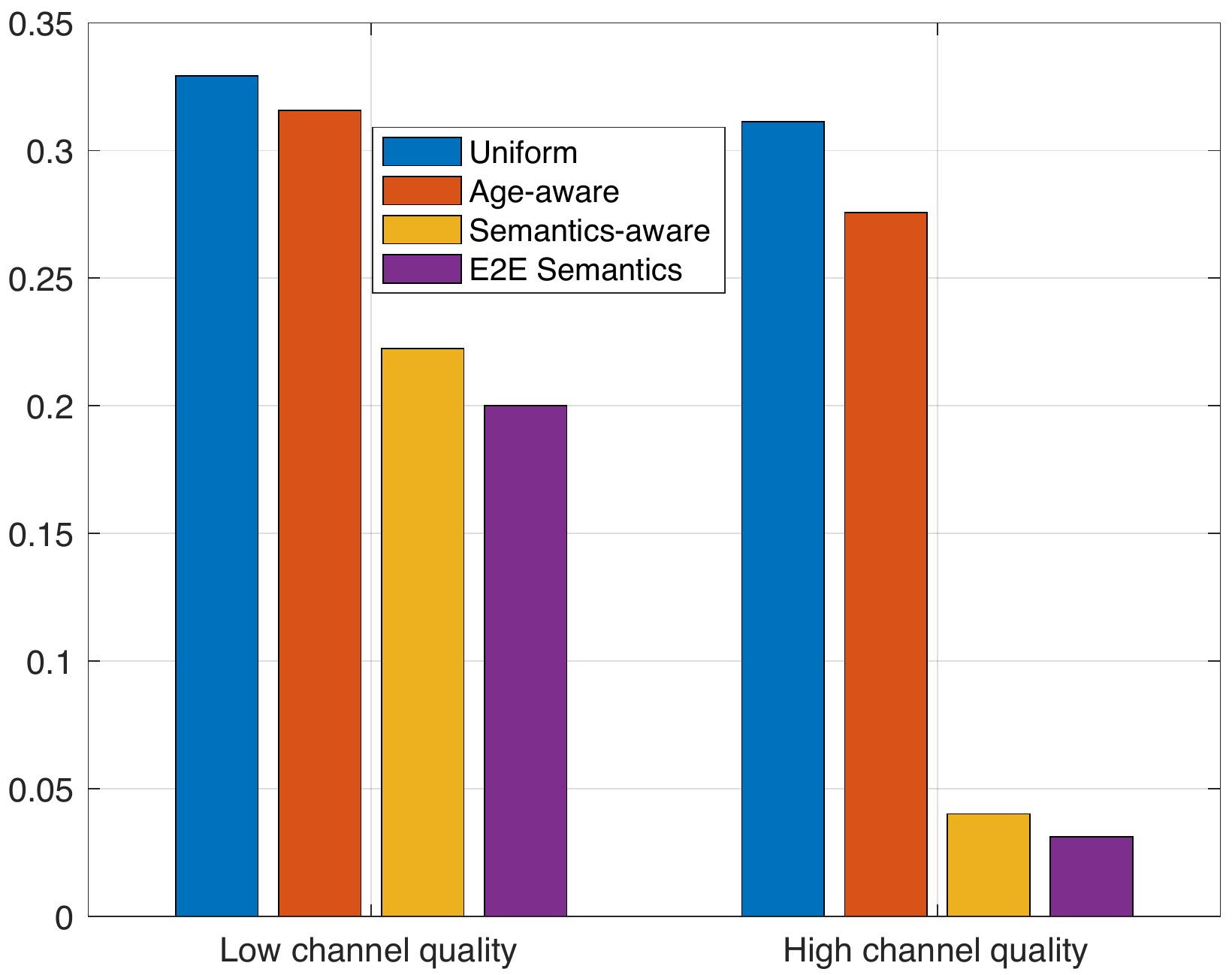}
		\caption{Real-time reconstruction error} 
		\label{fig:error_high_var}
	\end{subfigure}
	\vspace{1em} 
	\begin{subfigure}{0.32\textwidth} 
		\includegraphics[width=\textwidth]{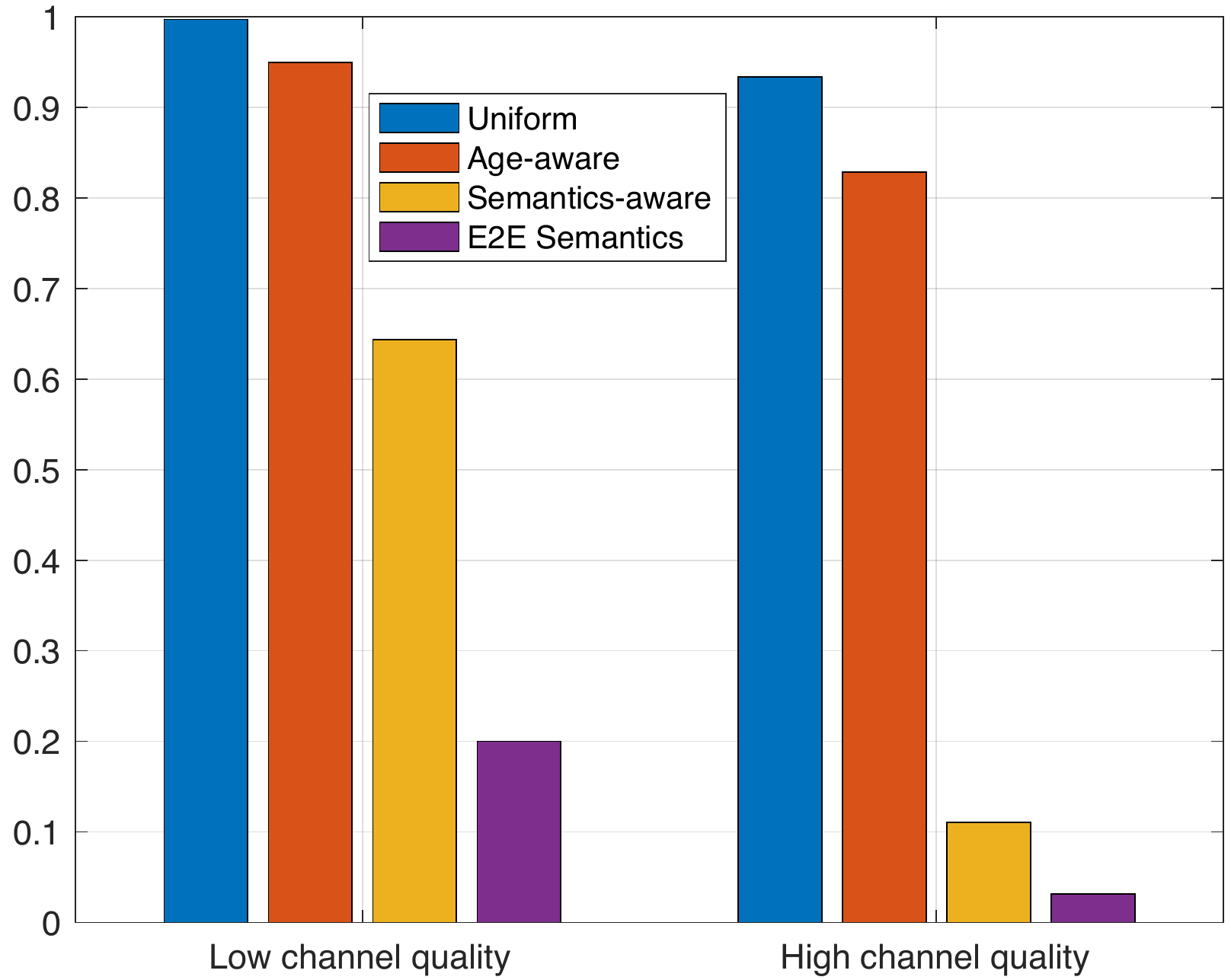}
		\caption{Cost of actuation error} 
		\label{fig:costacterror_high_var}
	\end{subfigure}\vspace{-0.2in}
	\caption{The case of a rapidly changing source ith $p=0.8$ and $q=0.3$.} 
	\label{fig:high_var}
\end{figure}

We have two scenarios regarding the source variability; the first being when the source is slowly changing, ($p=0.95$, $q=0.9$) 
depicted in Fig. \ref{fig:low_var}, and the second being when the source is rapidly changing, ($p=0.8$, $q=0.3$), 
depicted in Fig. \ref{fig:high_var}.
We observe that in the case of low source variability, the age-aware policy outperforms the semantics-aware one when the communication channel quality is poor. This is due to the fact that if a transmission error occurs and the receiver fails to anticipate the right state, this leads to high reconstruction error. This is the case where a uniform sampling policy could perform better. However, the performance from the perspective of cost of actuation error is different; the semantics-aware policy outperforms uniform sampling.

For slow varying sources, E2E semantics significantly outperforms the semantics-aware scheme due to the fact that the system manages to eliminate the discrepancy fast, even in the case of a low quality channel. 
On the other hand, for rapidly varying sources, both semantics-empowered policies exhibit similar reconstruction error performance, while E2E semantics provides the lowest actuation error without wasting resources transmitting uninformative samples. 

\textbf{Uninformative samples reduction.} We provide here the percentage of uninformative samples each policy generates. E2E semantics, by definition, does not create redundant samples, since it accounts for the end-to-end discrepancy among sources in both ends. Uniform and age-aware policies generate, in most cases, the highest percentage of redundant (uninformative) samples. Despite being expected for uniform sampling, the explanation for the age-aware scheme is as follows. 
For the purpose of real-time remote reconstruction, metrics based solely on information freshness are inefficient since baseline AoI does not take into account the source variability.  
The performance of semantics-aware policy is relatively good, despite operating only at the transmitter side. The percentage of uninformative samples is less than $10\%$ when the channel is good for a slow varying source and less than $5\%$ for a rapidly varying source.

In a nutshell, semantics-empowered policies basically generate \textit{informative samples}, i.e., samples conveying the most valuable information for the purpose of real-time reconstruction and actuation, for which the timing when this information has been acquired is crucial. 
Note that additional gains, mainly in terms of savings in communication load and in the number of samples generated, could be achieved by learning the patterns of the source evolution, for instance via reinforcement learning.

%Error and Actuation error for high variability source

%========================================================================================================%
%========================================================================================================%
\section{Semantics-Aware Networking}
In this section, we present key functionalities required for reliable and timely communication of concisely represented valuable information in semantics-aware networks. The main operations span from local goal-oriented information acquisition, representation, and semantic value inference, up to data prioritization mechanisms and in-network processing (e.g., fusion, compression). Specifically, they will allow to perform: 

%\begin{itemize}[noitemsep,topsep=0pt,partopsep=0pt]
\begin{itemize}[leftmargin=*,align=left]
\item \textit{\textbf{Semantic filtering}} for avoiding unnecessary redundancy during data acquisition and information encoding using active sampling and censoring. That way, only useful and relevant information is generated and transmitted. Moreover, semantics-aware data acquisition can significantly reduce the average sampling rate (sub Nyquist limit) and channel utilization without affecting reconstruction accuracy. 
\item \textit{\textbf{Semantic preprocessing}}, which enables goal-oriented sparse representations (e.g., feature extraction, labeling, embedding, segmentation) and computations on an information manifold. For example, a robot could compute local estimates of the state (tracked target's velocity and location) from visual features or scene labeling extracted from an image instead of sending raw data. Another instance is in distributed learning, where only data samples that are semantically representative (core set selection) or informative are processed and transmitted.
\item \textit{\textbf{Semantic reception}} for fast partial or approximate source reconstruction, as well as goal-dependent information recovery, fusion, and querying. Reconstruction quality is conventionally measured by distortion metrics, such as the time-averaged mean squared error (MSE) between the estimated $\hat{X}_t$ and the original input signal $X_t$, i.e., $\delta(X,\hat{X})=\frac{1}{T}\mathbb{E}\left[\int_0^T\left(X_t - \hat{X}_t\right)^2 \right]$, or via entropy-based measures. Depending on end-user's objectives, approximate results with different distortion or perceptual quality could be sufficient for achieving a specific goal. For example, a low quality, highly compressed video may be sufficient for remote surveillance during non-alert mode. 
Furthermore, in many scenarios involving, among others, images, patterns, and machine learning, low distortion does not necessarily mean high perceptual quality. In that case, reconstruction efficiency could be assessed using a {\emph{semantic quality indicator}} based on divergence measures $\mathcal{D}(p_X || p_{\hat{X}})$ or distance functions $d_X:\mathcal{X}\times\mathcal{X} \to \mathbb{R}_+$ on a sample space $\mathcal{X}$ (e.g., Kantorovich-Wasserstein), where $X$ ($\hat{X}$) is the input (output) signal with distribution $p_X$ ($p_{\hat{X}}$). In the illustrative example (Section~\ref{sec:example}), we saw that a scheme can have higher cost of actuation error (low perception quality) in spite of achieving lower reconstruction error (low distortion) than others.
\item \textit{\textbf{Semantic control}}, which enables agile orchestration of multi-quality multimodal information gathering and fusion and efficient resource utilization. Metadata processing is required for scalability whenever the amount of metadata carrying information on semantic attributes and the number of network nodes become prohibitively large. 
\end{itemize}

%========================================================================================================%
%========================================================================================================%

\section{Future Challenges}\label{sec:challenge}
We discuss now key open problems and technical challenges associated with this promising avenue of research.

\textbf{Semantic metrics:} A key challenge is to establish concrete metrics, which incorporate qualitative attributes of information in the existing communication theoretic edifice. These new semantics-based metrics should capture both source and network dynamics, as well as potential non-trivial interdependencies among information attributes.

\textbf{Semantics-aware Multiple Access:} Consider a large number of heterogeneous devices that transmit, either in time-, or event-triggered process-aware manner, signals conveying multi-quality information (not necessarily from the same codebook) to a remote destination. For optimally utilizing the shared medium, devices have to adapt their access patterns and transmission attempts not only based on exogenous traffic arrivals and the other nodes’ status, but also based on the source or process variability, the information semantics, and the applications' demands.

\textbf{Goal-oriented Resource Orchestration:} Semantic real-time data networking requires efficient scheduling and resource allocation policies for gathering multi-source multimodal - often correlated - information, acquired at different levels of quality. The objectives of emerging networked applications could be achieved by utilizing one of multiple alternative sets of multi-quality data objects. For example, a remote monitoring system could normally operate in non-alert mode using sensing data (e.g., images) with precision above $\alpha \%$ and freshness above $\beta$. These problems fall in the realm of real-time scheduling with multiple choices, for which online algorithms may select \textit{which} piece of information, from \textit{where} and \textit{when}, to gather and transmit under communication and processing resources constraints. 

\textbf{Multi-objective Stochastic Optimization:} Semantics-aware data gathering and prioritization require multi-criteria optimization with goal-oriented, end-user perceived utilities, which assess the relative degree of priority among different information attributes. A multi-objective stochastic optimization framework based on cumulative prospect theory \cite{CPT}, which incorporates semantics via risk-sensitive measures and multi-attribute entropy-based utility functions, and which performs rank-dependent nonlinear semantic weighting, seems to be a promising endeavor. 
%========================================================================================================%
%========================================================================================================%
\section{Epilogue}\label{sec:ccl} 
Supporting autonomous, real-time, and connected intelligence applications in future wireless networks necessitates fundamental theoretical advances in communication, information theory, and signal processing. It requires transforming commonly held design assumptions and prevailing communication paradigms. 
We proposed a structurally new, synergistic approach that accounts for the information semantics and aims at harnessing the high potential benefits of a goal-oriented unification of information generation, transmission, and usage, which have hitherto been treated in separation. 
Semantic networking will enable carrying around only the “most informative” data samples, thus conveying to the end user only information that is timely, useful, and valuable for achieving its goals. 
Semantics-empowered communication will significantly improve network resource usage, energy consumption, and computational efficiency, thus supporting the scalability of future massive, networked intelligent systems. 
It will pave the way for the design of next-generation real-time data networking and will provide the foundational technology for a plethora of socially useful services, including autonomous transportation, consumer robotics, environmental monitoring, and telehealth. 

\section*{Acknowledgment}
The authors would like to thank Anthony Ephremides and Petar Popovski for fruitful discussions. The second author acknowledges Elif Uysal and Onur Kaya for their collaborations in writing joint proposals on semantic-aware communication, which contributed to the vision and some of the material presented in this paper. The work of the second author has been supported by the Center for Industrial Information Technology (CENIIT), and the Excellence Center at Linköping-Lund in Information Technology (ELLIIT).

\bibliographystyle{IEEEtran}
\bibliography{Refs}
\end{document}